\newif\ifdoubleblind
\newif\iflackofspace
    \documentclass[sigconf]{acmart}

\usepackage[utf8]{inputenc}
\usepackage{amsmath,amsfonts}

\usepackage{algorithm}
\usepackage{listings}
\usepackage{algorithmic}
\usepackage{textcomp}
\usepackage{xcolor}
\usepackage[capitalise,noabbrev]{cleveref}
\usepackage{hyperref}

\usepackage{tabularx}
\usepackage{booktabs}

\AtBeginDocument{%
  }

\copyrightyear{2026}
\acmYear{2026}
\setcopyright{cc}
\setcctype{by}
\acmConference[ICPP Workshops '26]{Workshop Proceedings of the 55th International Conference on Parallel Processing}{September 28-October 01, 2026}{Singapore, Singapore}
\acmBooktitle{Workshop Proceedings of the 55th International Conference on Parallel Processing (ICPP Workshops '26), September 28-October 01, 2026, Singapore, Singapore}
\acmDOI{10.1145/3816891.3834891}
\acmISBN{979-8-4007-2763-4/2026/09}

\begin{document}

\title[Portable to Efficient: Auto-Tuning Hardware-Agnostic GPU Kernels in Julia]{Portable to Efficient:\\Auto-Tuning Hardware-Agnostic GPU Kernels in Julia}

\author{Floris-Jan Willemsen}
\email{f.q.willemsen@liacs.leidenuniv.nl}
\orcid{0000-0003-2295-8263}
\affiliation{%
  \institution{LIACS, Leiden University}
  \city{Leiden}
  \country{the Netherlands}
}

\author{Evelyne Ringoot}
\email{eringoot@mit.edu}
\orcid{0000-0003-4661-1483}
\affiliation{%
  \institution{Massachusetts Institute of Technology}
  \city{Cambridge}
  \state{Massachusetts}
  \country{USA}
}

\author{Alan Edelman}
\email{edelman@mit.edu}
\orcid{0000-0001-7676-3133}
\affiliation{%
  \institution{Massachusetts Institute of Technology}
  \city{Cambridge}
  \state{Massachusetts}
  \country{USA}
}

\renewcommand{\shortauthors}{Willemsen et al.}

\begin{abstract}

Traditionally, GPU kernels have been developed and optimized within vendor-specific programming models to achieve high performance, resulting in software that is difficult to optimize and adapt across increasingly heterogeneous computing systems. 
Hardware-agnostic programming models offer a more sustainable approach to GPU software development by improving portability and maintainability, but achieving efficient execution across diverse architectures remains challenging. 
We address this challenge by integrating auto-tuning into hardware-agnostic GPU kernels written in Julia. 

We rebuild the established \emph{Kernel Tuner} auto-tuning framework with Julia support, 
enabling systematic exploration of kernel configurations for hardware-agnostic GPU kernels targeting NVIDIA, AMD, Intel, and Apple GPUs. 
We demonstrate this approach on hardware-agnostic singular value decomposition (SVD) as implemented in the \emph{NextLA.jl} linear algebra library.

The results show that auto-tuning is essential for creating resource-efficient hardware-agnostic GPU kernels across a variety of hardware. 
Optimal configurations improve kernel performance by a factor of 3$\times$ to 7$\times$ compared to median parameter configurations, demonstrating the substantial impact of tuning on efficient hardware utilization.  %
Moreover, auto-tuning enables hardware-agnostic kernels to surpass manually optimized implementations in efficiency while reducing developer effort. %
This is made widely accessible through the first auto-tuning framework for Julia, \href{https://github.com/KernelTuner/KernelTuner.jl}{\emph{KernelTuner.jl}},
bringing portable yet resource-efficient GPU kernels within reach for a wide range of applications and devices.

\end{abstract}

\ccsdesc[500]{General and reference~Performance}
\ccsdesc[300]{Software and its engineering~Software performance}
\ccsdesc[100]{Computing methodologies~Parallel programming languages}

\keywords{Auto-tuning, Julia Language, GPU, HPC, Performance Optimization, Resource Efficiency, Dense Linear Algebra, Portability}  %

\maketitle

\section{Introduction}
\label{sec:introduction}

Modern HPC systems employ accelerators from multiple vendors and generations, requiring software to be efficient across diverse hardware platforms. %
Producing highly specific, optimized implementations for different architectures and vendors is costly and difficult to sustain, both in terms of development effort and long-term software maintenance. 
As HPC and workloads continue to evolve, improving the portability and efficiency of GPU software has become an important aspect of sustainable computing. 

Automatic performance tuning, or auto-tuning~\cite{balaprakashAutotuningHighPerformanceComputing2018}, has become a key technique in high-performance computing (HPC), enabling applications to achieve high efficiency across diverse hardware architectures without requiring manual, architecture-specific optimization. 
By systematically exploring large design spaces of functionally equivalent program variants, spanning thread configurations, tiling strategies, memory usage, and other parameters, auto-tuning adapts software to the characteristics of a target device and workload~\cite{hijmaOptimizationTechniquesGPU2023, vanwerkhovenLessonsLearnedDecade2020}. 
This approach is particularly important for GPU programming, where performance is highly sensitive to hardware-specific details and small changes in a configuration can result in large performance differences. 

Traditionally, achieving high performance on GPUs has relied heavily on vendor-specific programming models such as CUDA, and vendor-specific libraries such as CUBLAS/CUSOLVER for NVIDIA devices and ROCBLAS/ROCSOLVER for AMD devices in the case of linear algebra. 
Applications and libraries are then built based on these low-level kernels, limiting their expansion to novel hardware types. %
In various cases, this results in complete redevelopment to achieve optimal performance, such as for linear algebra libraries~\cite{deakin2019performance}. 
The increasing diversity in hardware and heterogeneity of computing clusters have inspired a range of innovations in programming abstractions to address these challenges~\cite{10820782, 9652859,10196600} by providing hardware-agnostic interfaces to program GPU kernels.
In particular, the Julia ecosystem has emerged as a promising platform for sustainable portable computing, offering high-level abstractions while retaining low-level control over execution. 
The \emph{NextLA.jl} library~\cite{the-code} builds on this vision by providing hardware-agnostic linear algebra kernels that target multiple GPU backends, including NVIDIA, AMD, Intel, and Apple Metal~\cite{ringoot2025performant, ringoot2026acceleratingbidiagonalizationbandedmatrices, carrica2025toward, dla}. 

However, achieving high performance across such diverse architectures is only possible by abstracting hardware-dependent variables in the low-level GPU kernels for each device and data type to exploit the characteristics of each hardware type maximally. This has traditionally been done by selection by expert knowledge or through a manually implemented brute-force search of selected parameters~\cite{ringoot2025performant, ringoot2026acceleratingbidiagonalizationbandedmatrices}, limiting the scale of optimizations as well as requiring additional development effort. 
Automating the tuning of GPU kernel parameters and making it widely accessible for users is, therefore, a critical component for realizing true sustainability, both in resource efficiency and development effort.
Existing auto-tuning frameworks provide powerful mechanisms for exploring tunable parameter spaces and optimizing performance~\cite{nugterenCLTuneGenericAutotuner2015, raschATFGenericAutoTuning2017, petrovicKernelTuningToolkit2023, vanwerkhovenKernelTunerSearchoptimizing2019}. 
However, these frameworks have primarily been designed for CUDA and similar vendor-specific programming models and have, to the extent of our knowledge, not been applied to fully hardware-agnostic frameworks, limiting their applicability in modern high-level, wide-target environments like Julia. 
Bridging this gap is essential to achieve true portability, where a single implementation of, e.g., linear algebra can achieve optimal performance across a range of consumer and High-Performance Computing GPU hardware, bringing full resource-efficient performance to hardware-agnostic programming models. 

In this work, we extend the established Kernel Tuner framework~\cite{vanwerkhovenKernelTunerSearchoptimizing2019} to support auto-tuning of Julia GPU kernels and provide it as the Julia package \href{https://github.com/KernelTuner/KernelTuner.jl}{\emph{KernelTuner.jl}}\footnote{\url{https://github.com/KernelTuner/KernelTuner.jl}}, enabling seamless integration with hardware-agnostic kernel definitions. 
We apply this to a kernel of a singular value decomposition (SVD) library implemented in Julia, demonstrating how auto-tuning can efficiently optimize performance across multiple GPU architectures. 
We show that auto-tuning achieves superior performance while reducing development effort. 

In summary, this work demonstrates that combining hardware-agnostic kernel design with automated tuning is a practical and effective step toward democratizing high-performance GPU computing across vendors. 
Our contributions are as follows:
\begin{enumerate}
    \item \textbf{Julia auto-tuning support.} We introduce the first auto-tuning framework for Julia GPU kernels as an accessible Julia \href{https://github.com/KernelTuner/KernelTuner.jl}{package, \emph{KernelTuner.jl}\footnotemark[1]}. 
    This enables systematic auto-tuning of hardware-agnostic kernels to target a wide variety of platforms, such as NVIDIA, AMD, Intel, and Apple. 
    \item \textbf{Cross-language integration.} We present a seamless integration of the Python-based \emph{Kernel Tuner} auto-tuning framework with a Julia-based hardware-agnostic GPU programming model that avoids duplicate codebases while providing mature optimization tooling. This design enables flexible mixed-language workflows, allowing Julia-based scripts to tune kernels written in any language supported by Kernel Tuner, while also enabling Python-based tuning scripts to operate directly on Julia kernels. 
    \item \textbf{Unlocking performance gains.} As auto-tuning allows searching a wider range of tunable parameters in a user-friendly and time-efficient manner, search spaces can be substantially expanded. 
    We demonstrate the effectiveness of this approach on a GPU kernel in the \emph{NextLA} linear algebra library, achieving performance improvements of 3$\times$ to 7$\times$ over median parameter configurations. 
\end{enumerate}

This novel approach bridges the performance gap between highly optimized vendor-specific implementations and hardware-agnostic kernels, enabling portable code to achieve optimal efficiency. 
By reducing the reliance on vendor-specific implementations, it not only improves application portability but also promotes fairer evaluation, ensuring that performance comparisons are not exclusive or biased by unequal levels of manual optimization across hardware platforms.

\section{Related Work} 
\label{sec:related_work}

Achieving high performance across diverse GPU architectures remains a longstanding challenge in high-performance computing. 
Motivated by the emergence of heterogeneous computing clusters~\cite{10820782, 9652859,10196600} and the resulting development and optimization burden, exemplified by multiple low-level linear algebra libraries in application libraries~\cite{10883176, deakin2019performance}, several programming frameworks have been proposed to enable hardware-agnostic GPU development. 
Examples of such efforts include MOJO~\cite{10.1145/3731599.3767573}, Kokkos~\cite{edwards2014kokkos}, SYCL, RAJA~\cite{beckingsale2019raja}, HIP~\cite{https://doi.org/10.1002/cpe.7866}, Triton~\cite{tillet2019triton}, and Julia KernelAbstractions~\cite{kajl}. 
Of these, the Julia GPU framework is of particular interest as it provides various levels of control of GPUs and targets not only the traditional GPU vendors NVIDIA and AMD, but also the more novel Intel and Apple GPUs~\cite{besard2019}. 

Traditionally, achieving high GPU performance across platforms has consisted of writing and optimizing GPU kernels on one single type of hardware, then translating them to other platforms and 
going through the same process again, which requires extensive architecture-specific rewrites~\cite{pennycook2016metric, deakin2019performance}. 
Instead, recent work has demonstrated that by writing GPU kernels in a tunable fashion~\cite{ringoot2025performant,reguly2023evaluating, davis2024evaluative}, true performance portability in line with vendor-optimized solutions is possible, but only through careful tuning of the parameters. 
In Julia, the \texttt{AcceleratedKernels}~\cite{nicusan2025acceleratedkernelsjlcrossarchitectureparallelalgorithms} and \texttt{NextLA}~\cite{dla} libraries provide such portable libraries for GPUs. 
In addition, research in automatic conversion of vendor-specific kernels to Julia's Kernel Abstractions facilitates accessible porting of kernels~\cite{de2024juliana}. 

In practice, such tunable parameter selection is executed through expert knowledge or brute-force searches, introducing several limitations: excessive use of computing resources, artificial restrictions of the search space, and additional development time for the inclusion of novel hardware. 
Auto-tuning has emerged as a key technique to address the challenge of adapting software configurations to the underlying hardware, efficiently exploring parameter spaces that commonly control aspects such as thread mapping, tiling, loop transformations, and memory usage~\cite{balaprakashAutotuningHighPerformanceComputing2018}. 
A wide range of auto-tuning frameworks have been proposed, including ATLAS~\cite{whaleyAutomatedEmpiricalOptimizations2001}, CLTune~\cite{nugterenCLTuneGenericAutotuner2015}, OpenTuner~\cite{anselOpentunerExtensibleFramework2014}, Kernel Tuner~\cite{vanwerkhovenKernelTunerSearchoptimizing2019}, and KTT~\cite{petrovicKernelTuningToolkit2023}. 
These systems enable systematic exploration of kernel configurations and can yield substantial performance improvements. %
Importantly, prior work has shown that tuning is not merely beneficial but often necessary~\cite{ryooProgramOptimizationSpace2008}, increasingly so on modern GPU architectures~\cite{lurati2024Bringing} where efficiency depends on complex interactions between parameters and hardware characteristics, reinforcing the need for automated tuning approaches.

Despite advances in both areas, the integration of auto-tuning with hardware-agnostic GPU programming models remains limited. Many existing auto-tuning frameworks focus on low-level languages and vendor-specific toolchains, while portable methods often rely on static heuristics or manual adjustment if at all possible. 
This gap suggests an opportunity to combine the strengths of both approaches and make true hardware-optimal resource efficiency for portable implementations possible: using auto-tuning to systematically optimize hardware-agnostic kernels across architectures, from consumer hardware to HPC GPUs.

\section{Background}
\label{sec:background}
This section provides a general introduction to the auto-tuning framework Kernel Tuner (\cref{subsec:background-kernel-tuner}) and the Julia GPU ecosystem (\cref{subsec:background-julia-gpu}).

\subsection{Kernel Tuner}
\label{subsec:background-kernel-tuner}

Kernel Tuner is a Python-based, open-source auto-tuning framework designed for optimizing computational kernels~\cite{vanwerkhovenKernelTunerSearchoptimizing2019}. It supports multiple programming backends, including CUDA, HIP, OpenCL, and OpenACC, C++ and Fortran, and is widely adopted for tuning applications across different architectures and objectives, such as performance, accuracy and energy efficiency~\cite{schoonhovenGoingGreenOptimizing2022}.

\begin{figure}[htb]
    \centering
    \includegraphics[width=0.95\linewidth]{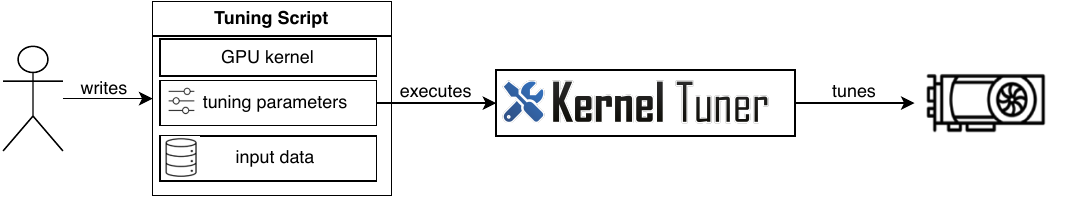}
    \vspace{-0.3cm}
    \caption{Basic overview of Kernel Tuner usage.}
    \Description[Kernel Tuner usage diagram]{Basic overview of how to use Kernel Tuner}
    \label{fig:kt-usage}
\end{figure}

\Cref{fig:kt-usage} shows how Kernel Tuner can be used. 
Users provide Kernel Tuner with a kernel implementation and a script that specifies the input data, a description of the tunable parameters, and possibly any constraints. At the start of the tuning process, Kernel Tuner constructs a search space $\mathcal{X}$ consisting of all valid configurations. The optimization algorithm, called an optimization strategy in Kernel Tuner, iteratively selects candidate solutions $x \in \mathcal{X}$ to be compiled, executed, and measured. 
The evaluation metric determines the quality of the configuration. 
Kernel Tuner internally abstracts the different backends into a unified interface, allowing most of the tool to operate independently of the programming language and target hardware platform. 

More formally, auto-tuning involves optimizing an application or {\em kernel} $K_i$ on a target system $G_j$ for input data $I_k$ on objective function $f_{G_j,I_k}(K_i)$. 
The auto-tuner constructs a search space $\mathcal{X}$ by considering all tunable parameters and their valid values, subject to user-defined constraints~\cite{willemsenEfficientConstructionLarge2025}. 
The objective is to determine the optimal configuration as follows (assuming minimization): 
\begin{equation} \label{eq:autotuning}
x^\star = \underset{x\in\mathcal{X}}{\text{arg min}} \, f_{G_j,I_k}(K_{i,x}).
\end{equation}

As the search space $\mathcal{X}$ can be very large, Kernel Tuner has a wide range of optimization algorithms which have been designed and tuned specifically for efficiently finding near-optimal configurations in auto-tuning problems~\cite{willemsenBayesianOptimizationAutotuning2021,willemsenTuningTunerIntroducing2025,willemsenAutomatedAlgorithmDesign2026}. 
In addition, it provides a variety of additional features such as verification checks, observers, accuracy tooling, and energy tuning~\cite{powersensor3,schoonhovenGoingGreenOptimizing2022}. 

\subsection{Julia GPU ecosystem}
\label{subsec:background-julia-gpu}
The Julia GPU ecosystem provides a high-level abstraction that is well-suited to portability: modularity and extensibility are integrated through multiple-dispatch, which allows overloading new methods to existing functions in a direct manner~\cite{bezanson2017julia}. 
The LLVM Just-In-Time compiler directly translates abstract code bases across data precisions and hardware to machine code through type inference~\cite{10.1145/3276490}, optimized for the specific data precision and hardware. 
The integration of advanced optimization features such as specialization, interoperability, loop unrolling, and vectorization enables achieving performance close to vendor-specific code~\cite{giordano2022productivity}. 
The GPU ecosystem includes hardware integrations of the main GPU vendors, NVIDIA ({\tt CUDA.jl}~\cite{besard2019}) and AMD ({\tt AMDGPU.jl}~\cite{AMDGPU}), as well as Intel ({\tt oneAPI.jl}~\cite{oneapi}) and Apple ({\tt Metal.jl}~\cite{metal})). 
Julia's \texttt{KernelAbstractions} then provides an interface targeting all of these platforms (\cref{subsec:kernel_abstractions}), which the NextLA library leverages to construct a fully hardware-agnostic yet performant GPU linear algebra library (\cref{subsec:nextla}). 
The integration of the Kernel Tuner framework is achieved through Julia-Python interfaces (\cref{subsec:juliapython}).

\subsubsection{KernelAbstractions}
\label{subsec:kernel_abstractions}
The {\tt GPUArrays} abstraction~\cite{gpuajl, besard2019} provides an interface for high-performance computing on GPUs that targets a wide range of hardware backends. 
\emph{KernelAbstractions} is a specialized interface providing access to kernel functionality across hardware vendors, compiling the generic kernel interface written for GPUArrays to hardware-specific Intermediate Representations at compile time. 
It provides direct access to all GPU memory levels, similar to a CUDA kernel. 
Earlier work leveraging {\tt KernelAbstractions.jl} demonstrated the viability of the approach for scientific machine learning~\cite{UTKARSH2024116591}. KernelAbstraction allows passing certain compile-time constants as function variables, a feature allowing compiling the same kernel for multiple such constants inside a single Julia module, which substantially facilitates auto-tuning.

\subsubsection{NextLA}
\label{subsec:nextla} NextLA~\cite{dla} is an abstraction-based linear algebra library implementing generic Julia functions for dense numerical linear algebra. It provides three types of abstraction layers:
\begin{enumerate}
    \item Abstraction across data precision and CPU-GPU hardware, through abstract high-level implementations calling into libraries for low-level functions such as matrix multiplication~\cite{dla}.
    \item Low-level GPU function implementations for specific kernels that are not available in all low-level libraries or that underperform, such as for the singular value decomposition, written in KernelAbstractions, designed for portability through careful selection of tunable parameters~\cite{ringoot2025performant,ringoot2026acceleratingbidiagonalizationbandedmatrices}.
    \item Abstraction across hardware size, from single GPU, to out-of-core, to multi-GPU, through a single algorithmic interface where communication is included where needed~\cite{ringoot2025scaling}.
\end{enumerate}
The library provides future-proof, efficient, and generic alternatives to legacy libraries. 
The second type of abstractions provided by the NextLA library is where the inclusion of auto-tuning is crucial to achieve true sustainability: optimizing resource efficiency on any type of GPU hardware.

\subsubsection{Python-Julia integration}
\label{subsec:juliapython} 
In the HPC and scientific community, the debate on whether to use Python or Julia has been long-standing~\cite{10.1007/978-3-031-70906-7_4, 10196600, Eschle2023, Coleman2021, churavy2022bridging}, with both communities having developed strong ecosystems of libraries and users, and some packages only available in one language. 
Application developers frequently do not desire rewriting their existing library in a whole new language, and are limited by their initial choice. 
In this work, we achieve a symbiosis between the two programming languages, applying the Python-based Kernel Tuner framework to Julia with minimal development and user impact. 
We utilize the PythonCall~\cite{PythonCalljl} and CondaPkg~\cite{CondaPkgjl} packages to achieve this integration, benefiting from the existing Kernel Tuner in Python and the existing hardware-agnosticism in Julia for a symbiotic solution for a complete hardware-portability solution.

\section{Integration of Julia into Kernel Tuner}
\label{sec:design_implementation}
\begin{figure}[htb]
    \centering
    \includegraphics[width=0.98\linewidth]{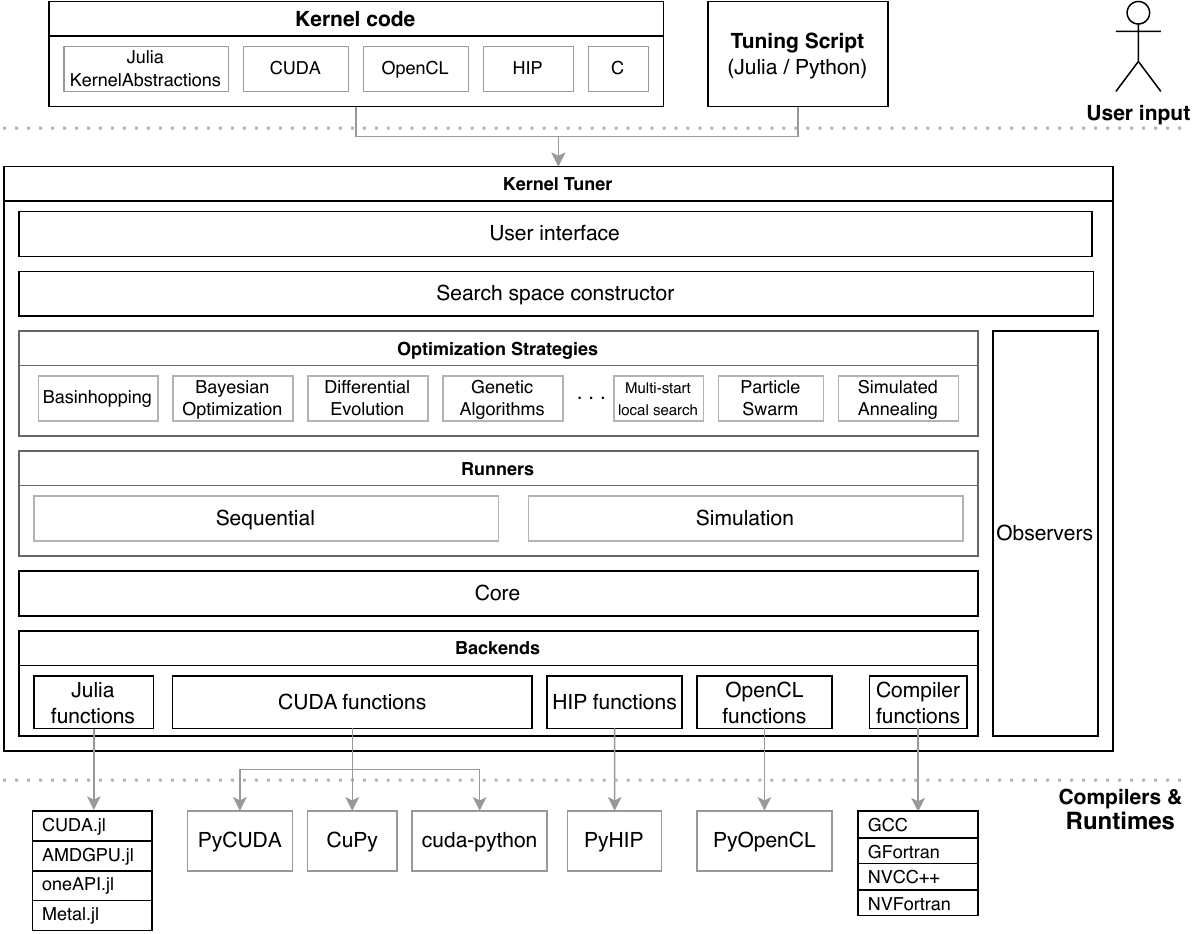}
    \vspace{-0.3cm}
    \caption{Overview of the Kernel Tuner software architecture.}
    \Description[Kernel Tuner software architecture]{Overview of the software architecture of Kernel Tuner}
    \label{fig:kt-arch}
\end{figure}

To enable auto-tuning of hardware-agnostic Julia GPU kernels, we extend the Kernel Tuner framework with a dedicated Julia backend and interface, as shown in \cref{fig:kt-arch}. 
This design bridges the gap between Python-based auto-tuning infrastructure and Julia’s GPU ecosystem, while preserving the flexibility and characteristics of both environments without duplication. 
Our implementation consists of two main components: (i) a Julia backend for Kernel Tuner described in \cref{subsec:design_implementation_julia_backend,subsec:design_implementation_timing_observers,subsec:design_implementation_julia_helper}, and (ii) a Julia-facing interface (\cref{subsec:design_implementation_julia_package}), allowing Kernel Tuner to be used natively from Julia.

\subsection{The Julia Backend}
\label{subsec:design_implementation_julia_backend}
The core of our implementation is a new backend that enables Kernel Tuner to compile, launch, and measure Julia GPU kernels. 
This backend leverages JuliaCall~\cite{PythonCalljl} to embed a Julia runtime within Python, allowing direct invocation of Julia code and seamless interoperability between the two languages. 

The backend extends the existing \texttt{GPUBackend} abstraction in Kernel Tuner and follows the standard workflow of compilation, memory management, kernel execution, and performance measurement. 
During initialization, the backend automatically determines available Julia GPU backends (i.e., CUDA, AMDGPU, oneAPI, and Metal) using system-level queries. 

Kernel compilation is performed by dynamically constructing a Julia module that wraps the user-provided kernel code. 
This approach isolates kernel definitions and avoids naming conflicts across tuning runs. 
Once compiled, the kernel is retrieved as a callable Julia function and stored for subsequent execution. 

Memory management is handled by converting Python host arrays into backend-specific Julia GPU arrays. %
This conversion is performed transparently through helper functions defined in a lightweight Julia module. 
Each argument is deep-copied before transfer to avoid unintended side effects during tuning. 
Device-to-host and host-to-device memory transfers are implemented using direct array copying. %

Kernel execution is implemented using Julia’s KernelAbstractions.jl (\cref{subsec:kernel_abstractions}), which provides a unified interface for launching GPU kernels across backends. For each configuration, Kernel Tuner passes the parameter values, grid size, and workgroup size to a Julia-native layer (\cref{subsec:design_implementation_julia_helper}) that constructs and launches the configured kernel. 
Tunable parameter values are converted to Julia \texttt{Val} types to enable compile-time specialization, ensuring that tuning parameters are fully exposed to the Julia compiler. 
Each evaluated configuration includes an initial warm-up execution that is excluded from the reported metrics, ensuring that Julia’s just-in-time (JIT) compilation is triggered and does not affect the measured kernel performance. 

\subsection{Cross-Backend Timing and Observability}
\label{subsec:design_implementation_timing_observers}
Accurate performance measurement across heterogeneous backends is a key challenge. To address this, we introduce a custom \texttt{JuliaRuntimeObserver} that integrates with Kernel Tuner’s Observer framework seen in \cref{fig:kt-arch}. 
This observer provides backend-specific timing and event mechanisms while maintaining a uniform interface.
For CUDA and AMD backends, we use GPU stream event timing (CuEvent and HIPEvent, respectively), enabling precise measurement of kernel execution time. For Metal, where event-based timing is less directly accessible, we measure execution time by inserting command buffers before and after kernel execution and computing the elapsed GPU time. For oneAPI backends, where event timing is currently unavailable, we fall back to host-side timing, with appropriate warnings about reduced accuracy.

To ensure robustness, the implementation includes safeguards against timing inconsistencies. %
Additionally, kernel execution is wrapped to capture and propagate output and exceptions, allowing Kernel Tuner to gracefully skip invalid configurations without interrupting the tuning process. 

\subsection{Julia Helper Layer}
\label{subsec:design_implementation_julia_helper}
A thin Julia helper layer provides the interaction between Kernel Tuner and KernelAbstractions. This layer mainly implements a generic kernel launcher that handles parameter passing, execution, synchronization, and timing. 
The launcher abstracts backend-specific details, including event recording and synchronization, and ensures that all kernels are executed within a consistent interface. 

Importantly, this design allows Kernel Tuner to treat Julia kernels similarly to CUDA or OpenCL kernels, minimizing changes to the core tuning logic and extensive supporting tooling while enabling support for multiple GPU backends through Julia. 

\subsection{The Julia Interface \& Package}
\label{subsec:design_implementation_julia_package}
To make Kernel Tuner accessible to Julia users, we provide a Julia interface using PythonCall~\cite{PythonCalljl}, which enables calling Python code from Julia with minimal overhead. This allows users to define kernels and tuning problems entirely in Julia, while transparently leveraging the extensive Kernel Tuner optimization, verification, and data collection capabilities. 

In this workflow, users write KernelAbstractions kernels in Julia and define tunable parameters and arguments using a combination of native Julia constructs, while also being able to use Kernel Tuner structures like \texttt{Tunable} and \texttt{OptAlgWrapper}~\cite{willemsenAutomatedAlgorithmDesign2026}. 
Kernel Tuner is invoked through PythonCall, passing the kernel source, parameter space, and execution configuration. 
The Julia backend ensures that all kernel compilation and execution occur within the Julia runtime, while Python handles the orchestration of the tuning process.

\ifdoubleblind
\else
This design enables a seamless user experience: developers can remain within the Julia ecosystem while benefiting from the mature auto-tuning infrastructure of Kernel Tuner, easily accessible as a \href{https://github.com/KernelTuner/KernelTuner.jl}{package}. 
\fi
At the same time, it avoids duplicate code bases for Kernel Tuner and preserves interoperability, and in addition allows mixed-language workflows where Julia-based scripts can tune kernels in any of the other languages and programming models compatible with Kernel Tuner, as well as having Python-based tuning scripts operate directly on Julia kernels.

\section{Evaluation}
\label{sec:evaluation}
In this section, we evaluate the advancements presented in \cref{sec:design_implementation} to determine their impact.
First, we discuss the application used for evaluation in \cref{subsec:evaluation_nextla}.
The experimental setup is described in \cref{subsec:evaluation_setup_hardware}. 
Finally, we discuss the evaluation results in \cref{sec:results}. 

\begin{table*}[htb]
    \centering
    \caption{Hardware and Software Specifications of the Systems used in Evaluation}
    \label{tab:hardware_software_specs}
    \footnotesize 
    \vspace{-0.3cm}
    \begin{tabularx}{\linewidth}{@{}lXXXXXX@{}}
        \toprule
        \textbf{System Name} & \textbf{GPU} & \textbf{VRAM} & \textbf{CPU} & \textbf{RAM} & \textbf{OS / Kernel} & \textbf{Software Stack} \\
        \midrule
        AMD University Program~\cite{amd_hpcfund_intro} & AMD MI300X & 192 GB HBM3 & EPYC 9684X 16-core & 256 GB & Rocky Linux 9.4 (Kernel 5.14) & ROCM 7.1.0, Julia 1.12.5 \\ \addlinespace
        AMD University Program~\cite{amd_hpcfund_intro} & AMD MI210 & 64GB HBM2e & EPYC 7V13 16-core & 64 GB & Rocky Linux 9.4 (Kernel 5.14) & ROCM 7.1.0, Julia 1.12.5 \\ 
        \addlinespace
        \midrule
        MIT Engaging~\cite{mit_orcd_systems} & Nvidia H200 & 141 GB HBM3e & Intel Xeon Platinum 8580 120-core & 2011 GB & Rocky Linux 8.10 (Kernel 5.14) & CUDA 13.1, Julia 1.12.3 \\ \addlinespace
        MIT Engaging~\cite{mit_orcd_systems} & Nvidia L40s &  48 GB GDDR6 & Intel Xeon Platinum 8562Y+ 64-core & 1005 GB & Rocky Linux 8.10 (Kernel 5.14) & CUDA 13.1, Julia 1.12.3  \\
        \addlinespace
        DAS-6~\cite{balMediumScaleDistributedSystem2016} & Nvidia A4000 & 16 GB GDDR6 & AMD EPYC-2 7402P 24-core & 128 GB & Rocky Linux 8 (Kernel 4.18) & CUDA 12.6, Julia 1.12.5 \\
        \addlinespace
        \midrule
        Stampede 3 & Intel Data Center GPU Max 1550s & 128 GB HBM2e & Intel Xeon Platinum 8480 96-core & 1 TB & Rocky Linux 9.7 (kernel 5.14) & oneAPI Level Zero 1.3, Julia 1.12.6 \\ \addlinespace
        \midrule
        Apple MacBook Pro 14" 2021 & M1 Pro 16-core & 32GB LPDDR5 (Unified) & M1 Pro 10-core & 32GB LPDDR5 (Unified) & macOS 26.4.1 & Metal 4, Julia 1.12.6 \\ \addlinespace
        \bottomrule
    \end{tabularx}
\end{table*}

\subsection{Performance evaluation in NextLA.jl} 
\label{subsec:evaluation_nextla}

We examine the benefits of Kernel Tuner on one of the most challenging GPU kernels: the calculation of singular values~\cite{ringoot2025performant, ringoot2026acceleratingbidiagonalizationbandedmatrices}. 
In particular, we focus on the first stage of the algorithm: the reduction to banded form and the GPU kernel for the compute-bound trailing matrix update.
\ifdoubleblind
\else
We summarize the algorithm in \cref{subsec:evaluation_setup_benchmark} and the GPU kernels with their parameters in \cref{subsect:gpukernels} for completeness; further details can be found in prior work~\cite{ringoot2026acceleratingbidiagonalizationbandedmatrices, ringoot2025performant}.
\fi

\subsubsection{Algorithm}
\label{subsec:evaluation_setup_benchmark}
\begin{figure}[ht]
    \centerline{\includegraphics[width=0.98\linewidth]{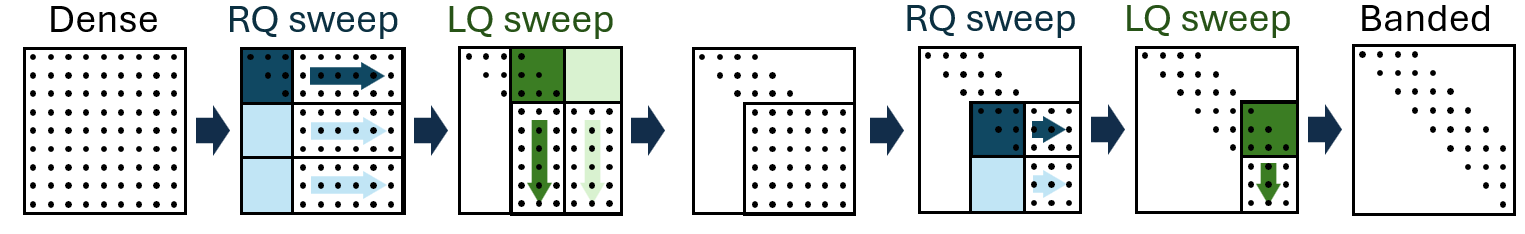}}
    \vspace{-0.3cm}
    \ifdoubleblind
\else
        \caption{The classic two-stage QR-based SVD, from~\cite{ringoot2025performant}.}
    \fi
    \label{fig:brdalgfig}
    \Description[Two-stage QR-based SVD.]{Overview of a classic two-stage QR-based Singular Value Decomposition (SVD).}
\end{figure}

The classic QR-based singular value algorithm consists of reducing the matrix to banded form and finally to bidiagonal and diagonal form, through the selection of appropriate orthogonal matrices that 'annihilate' or zero out a number of columns below the diagonal or rows above the diagonal. On GPUs, this is executed in tiles to improve data locality: for every diagonal tile of the matrix, the tiles below are annihilated, and the tiles to its right are as well, as shown in \cref{fig:brdalgfig}, until a banded matrix is achieved. 

\begin{algorithm}
\caption{Singular value calculation stage 1: reduction to band form, adapted from \cite{ringoot2025performant}}
\label{alg:redtoband}
\begin{algorithmic}[1]
\FOR{Diagonal tile $k$ : 1 $\rightarrow$ N }
\STATE{RQ sweep:}
\STATE{Fused GEQRT/TSQRT(TileColumn$_{k:end,k}$)} 
\STATE{Fused TSQRT/TSMQR(Trailing matrix$_{k:end,k:end}$)} 
\STATE{LQ sweep:}
\STATE{Fused GEQRT/TSQRT(TileRow$_{k,k+1:end}$)} 
\STATE{Fused TSQRT/TSMQR(Trailing matrix$_{k:end,k+1:end}$)} 
\ENDFOR
\end{algorithmic}
\end{algorithm}

Algorithm~\ref{alg:redtoband} details the first stage of the SVD: the reduction from dense to banded form. The GEQRT/TSQRT functions (the panel factorization) calculate such an orthogonal matrix, and are memory-throughput-bound functions with lower parallelism. The UNMQR/TSMQR functions (the trailing matrix update) then apply this orthogonal matrix to the remainder of the matrix to keep the singular values intact, and are compute-bound kernels with high degrees of parallelism.

 \subsubsection{GPU kernel description}\label{subsect:gpukernels}
We demonstrate the impact of the kernel tuning on the trailing matrix update kernel \textit{Fused TSQRT/TSMQR} that applies the panel factorization of a tile column or tile row to the trailing matrix, respectively, right or below it. Auto-tuning allows us to expand the parameters for the kernels further than was previously practical for a larger search space. The conceptual execution of the kernels is described in  Algorithm~\ref{alg:unmqr}.

\begin{algorithm}[ht]
\small
\caption{GPU kernel fused UNMQR/TSMQR: N thread blocks with COLPERBLOCK (CPB) threads \raggedright }\label{alg:unmqr}
\begin{algorithmic}[1]
 \renewcommand{\algorithmicrequire}{\textbf{INPUT}}
   \renewcommand{\algorithmicensure}{\textbf{Number of}}
    \REQUIRE \textbf{parameters} Tilesize TS, Tileheight TH, Split-k K
     \REQUIRE \textbf{parameters} Columns per block CPB
   \ENSURE \textbf{threads:} (CPB, K)
   \ENSURE \textbf{threadblocks:} N/CPB
 \REQUIRE \textbf{/ OUTPUT:}  A (N$\times$TS), X (N $\times$ N), $\hat{\tau}$ (TS)
\STATE{{\bf Thread memory}: $X_{i,j}$ (TH/K), $X0_{i,j}$ (TH/K)}
\STATE{{\bf Block memory}: $A_k$ (TH), $\hat{\tau}'$ (TS),  Cache$ (K,CPB)$  }
\FOR{Tile $t$:$1\rightarrow(N/TH)$}
\STATE{$\forall$ Thread $i,j$: $X_{i,j} \gets X[$tile t$,:] \qquad$ (or if $t=1$, load into $X0_{i,j}$)  }
\STATE{$\forall$ Threads $i,j$: $\hat{\tau}' \gets \hat{\tau}  \qquad$   }
\FOR{ $k$ : 1 $\rightarrow$ TS-1 }
\STATE{ $\forall$ Threads $i,j$: $A_k \gets A$[:,k] }
\STATE{synchronize threads}
\STATE{$\forall$ Thread $i\geq k$: cache$[i,j] \gets A_{i,j} \cdot X_{i,j}$ (vector product)}
\STATE{synchronize threads}
\STATE{$\forall$ Thread $i,j$: {$|A_i|^2$} $\gets  $sum(cache[:,i])}
\STATE{$\forall$ Thread $i,j$: calculate $X_{i,j}$, $X0_{i,j}$ }
\STATE{synchronize threads}
\ENDFOR
\STATE{$\forall$ Thread $i$ in block $n$: $ X$[tile t, :] $\gets X_{i,j}$   }
\ENDFOR
\STATE{$\forall$ Thread $i$ in block $n$: $ X$[tile 1, :] $\gets X0_{i,j}$ \qquad  }
\end{algorithmic}
\end{algorithm}

\subsection{Experimental setup}
\label{subsec:evaluation_setup_hardware}
To obtain a diverse set of real-world auto-tuning cases for evaluation, we use a diverse set of seven GPUs from four different vendors. 
The hardware and software specifications of each platform are detailed in \cref{tab:hardware_software_specs}. 

\begin{figure*}[htbp]
    \centering
\includegraphics[width=0.99\linewidth]{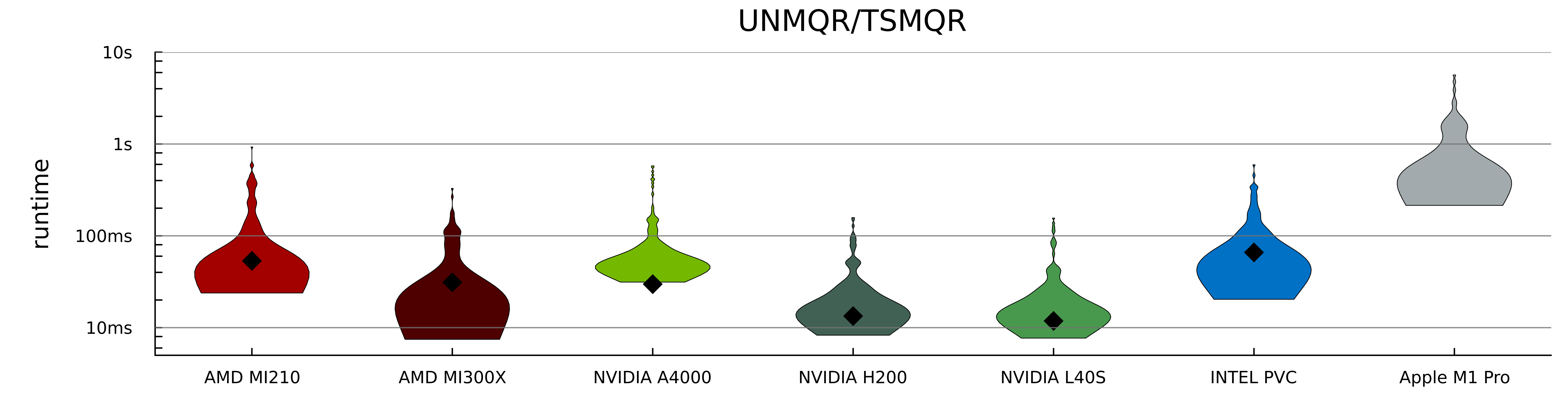}
    \vspace{-0.3cm}
    \caption{Runtime distribution of the GPU kernels over all kernel parameters for tilesize 64, by hardware type (left to right, AMD 210X, AMD MI300X, NVIDIA A4000, NVIDIA H200, NVIDIA L40S, Intel PVC Max, Metal M1). Black diamond markers indicate the originally selected kernel parameters in a reduced search space. 
    }
    \label{fig:kernels}
    \Description[Runtime violin plots per kernel]{Runtime distribution of the GPU kernels over all kernel parameters for tilesize 64, by hardware type (left to right, AMD 210X, AMD MI300X, NVIDIA A4000, NVIDIA H200, NVIDIA L40S, Intel PVC Max, Metal M1). Black diamond markers indicate the originally selected kernel parameters in a reduced search space. The distribution demonstrates the influence of increasing the searchable parameter space through user-friendly auto-tuning: in all but one case, a better configuration can be selected.}
\end{figure*}

\begin{table}[htbp]
    \centering
    \caption{Summary of optimal and median kernel runtimes across GPUs by kernel for tilesize 64.}
    \label{tab:medianoptimaltimes}
    \small
    \vspace{-0.3cm}
\begin{tabularx}{\linewidth}{p{1cm}|XXXXXXl}
\textbf{Runtime  } & \multicolumn{2}{l}{\textbf{AMD}} & \multicolumn{3}{l}{\textbf{NVIDIA}} & \textbf{INTEL} & \textbf{APPLE} \\
          (ms)              & MI210           & MI300X          & A4000      & H200       & L40S      & PVC            & M1 Pro            \\ \hline
  
Optimal                 & 5.7             & 1.8            & 10.5       & 2.9        & 4.5       & 9.1            & 80             \\
Median                  & 32              & 12             & 36         & 11         & 12        & 32             & 244            \\
Ratio                   & 5.6             & 6.7            & 3.4        & 3.8        & 2.7       & 3.5            & 3.1      \\ \hline          %

\end{tabularx}
\end{table}

Measurements on the MacBook Pro were performed under the following controlled conditions: a dedicated clean user profile with no background processes, sleep transitions deactivated, and the hardware restricted to AC power and the internal display to ensure performance stability. The runtime of each kernel is measured using stream event recording, where available (NVIDIA and AMD), and using command buffers and host-side timing if unavailable (in the case of Metal and Intel, respectively) as described in \cref{subsec:design_implementation_timing_observers}. 
To demonstrate the impact of auto-tuning, we focus on a specific matrix size and data precision, $16384$x$16384$ single-precision dense matrix, and evaluate the impact on its runtime of the tunable parameters across the search space.

\iflackofspace
We distinguish two types of tunable parameters: pipeline parameters, which change the number of kernels launched across the entire algorithm, and kernel parameters, which might change calculations inside the kernels, but do not influence how many consecutive kernels are launched until algorithm completion. 
\else
We distinguish between two types of tunable parameters: pipeline parameters, which change the number of kernels launched across the entire algorithm, and kernel parameters, which might change calculations inside the kernels, but do not influence how many consecutive kernels are launched until algorithm completion. In the first stage of the singular value decomposition, the tile size, which determines the number of diagonal tiles (Algorithm~\ref{alg:redtoband}), is the only pipeline parameter.
The runtime of the kernel can be directly compared across kernel parameters, but will naturally differ over pipeline parameters. The optimal combination of parameters across the entire SVD will then depend on the combination of several kernels.
\fi
In this experimental setup, we optimize the remaining individual GPU kernels across the search space of kernel parameters for each pipeline parameter value. Future work involves the extension to full pipeline optimization. The search space for the tunable parameters is as follows:
\begin{center}
    \centering
    \small
\begin{tabular}{r|l}
\hline 
Tilesize (TS)         & 8, 16, 32, 64, 128            \\ 
Tile height (TH)            & 8, 16, 32, 64, 128            \\ 
Split-K (K)          & 1, 2, 4, 8, 16, 32, 64, 128            \\
Columns per block (CPB)             & 1, 2, 4, 8, 16, 32, 64, 128 \\\hline
\end{tabular}
\end{center}

\subsection{Impact of auto-tuning}\label{sec:results}

\Cref{fig:kernels} demonstrates the impact on the runtime of individual kernels across their entire search space of parameters, for tile size 64. The mean and optimal values and their ratio are also shown in \cref{tab:medianoptimaltimes}. Tile size 64 is the parameter that was chosen during the initial SVD algorithm design through brute-force parameter search. Across all architectures, significant variation can be observed between the best-performing and median configurations, ranging from a factor of 2 to 7. The black diamond marker represents the configuration previously chosen when designing the SVD algorithm. 
\iflackofspace
\else
We observe in almost all cases the optimal parameter combination outperforms the expert-selected value, except for the A4000 GPU. This is expected when a library has been highly optimized for one architecture, and optimization for other architectures is more limited. This finding highlights the importance of auto-tuning for performant, sustainable portability.
\fi
Previously, manual tuning resulted in the artificial restriction of the search space based on expert opinion. Extending this to include a wider range of search parameters and values has enabled improving the kernel performance, as better-performing configurations are found in all but one case. 
No black diamond is shown for the Apple Metal GPU, as tile size 16 was originally selected for that hardware.

We find that, for this particular experiment, NVIDIA and AMD hardware outperform the Apple Metal M1 GPU. 
This is expected as the Apple M1 GPU is a consumer-grade laptop GPU, while the other hardware is designed for HPC settings. Additionally, Apple Metal GPUs feature unified memory, which changes the memory bandwidth and latency. However, the goal is not equal performance on all hardware, but rather optimal resource efficiency for the hardware at hand, irrespective of which hardware it is. 
With the integration of Kernel Tuner and Julia, GPU code developed and targeted on local hardware, such as a MacBook, can readily be scaled up to large-scale computing clusters, and vice versa.

\section{Conclusions}
\label{sec:conclusions}

In this work, we addressed the longstanding sustainability challenge of achieving both portability and high resource efficiency across diverse architectures by integrating Julia's hardware-agnostic GPU programming model into the established auto-tuning framework Kernel Tuner. 
By supporting Julia GPU kernels and providing the first Julia auto-tuning framework package, we enable systematic optimization of hardware-agnostic kernels without requiring architecture-specific rewrites. 
Moreover, we have demonstrated the value of a symbiotic approach between the Julia and Python ecosystems, where, through joining forces, novel approaches can be formulated with limited duplication. %

Through evaluation on singular value decomposition (SVD), we demonstrated that auto-tuning can effectively bridge the performance gap between portable implementations and vendor-specific optimizations. 
Our results show that automatically tuned configurations achieve superior performance compared to manually optimized approaches, while significantly reducing the developer effort and expertise required, 
enabling high efficiency on both established and emerging platforms.  
The resulting speedups of 3$\times$ to 7$\times$ relative to median parameter configurations highlight the necessity of combining high-level, hardware-agnostic programming models with automated performance optimization to obtain resource efficiency. 
Future work includes extension towards full pipeline tuning capability and Julia-specific energy efficiency observers. 

Overall, our work demonstrates that auto-tuning is a crucial enabler for sustainability in modern heterogeneous systems. 
By integrating auto-tuning into the Julia ecosystem, we take an important step toward democratizing high-performance computing, making near-optimal performance accessible across vendors without sacrificing programmability or maintainability. 

The auto-tuner presented in this work is available as the user-friendly open-source Julia package \href{https://github.com/KernelTuner/KernelTuner.jl}{KernelTuner.jl}\footnote{\url{https://github.com/KernelTuner/KernelTuner.jl}}.

\begin{acks}
\ifdoubleblind
\else
    We would like to extend our gratitude for the work and support of members of the Julia Lab, in particular Valentin Churavy and Rabab Alomairy. 
    The authors acknowledge the MIT Office of Research Computing and Data, Texas Advanced Computing Center (TACC) at the University of Texas at Austin (TG-CIS250776 through Advanced Cyberinfrastructure Coordination Ecosystem (ACCESS) program~\cite{ACCESS}, supported by U.S. NSF 2138259, 2138286, 2138307, 2137603, and 2138296), Advanced Micro Devices, Inc. under the AMD University Program’s AI \& HPC Cluster. 
    This material is based upon work supported by the US NSF (CNS-2346520, PHY-2028125, RISE-2425761, DMS-2325184, OAC-2103804, OSI-2029670), DARPA (HR00112490488), DoE (DE-NA0003965), NNSA (DE-NA0004266), USAFR (FA8750-19-2-1000). 
    The U.S. Government, its agencies and employees do not make any warranty, do not assume any liability or responsibility, do not make any endorsement for anything in this report, and do not represent that its use would not infringe privately owned rights. 
    The views and opinions of authors expressed herein are those of the authors alone. 
    E.R. was supported by a Fellowship from the MIT Office of Research Computing and Data. 
    This work is the result of a collaboration between the authors and Ben van Werkhoven, supported by the MIT General Seed Fund.
\fi
\end{acks}

\bibliographystyle{ACM-Reference-Format}
\bibliography{references_export}

\end{document}